# Investigating How Practitioners Use Human-AI Guidelines: A Case Study on the People + AI Guidebook

Nur Yildirim*
Carnegie Mellon University
Pittsburgh, USA
yildirim@cmu.edu

Mahima Pushkarna
Google Research
Toronto, Canada
mahimap@google.com

Nitesh Goyal
Google Research
New York, USA
teshg@google.com

Martin Wattenberg†
Google Research, Harvard University
Cambridge, USA
wattenberg@google.com

Fernanda Viégas†
Google Research, Harvard University
Cambridge, USA
viegas@google.com

## ABSTRACT

Artificial intelligence (AI) presents new challenges for the user experience (UX) of products and services. Recently, practitioner-facing resources and design guidelines have become available to ease some of these challenges. However, little research has investigated if and how these guidelines are used, and how they impact practice. In this paper, we investigated how industry practitioners use the People + AI Guidebook. We conducted interviews with 31 practitioners (i.e., designers, product managers) to understand how they use human-AI guidelines when designing AI-enabled products. Our findings revealed that practitioners use the guidebook not only for addressing AI's design challenges, but also for education, cross-functional communication, and for developing internal resources. We uncovered that practitioners desire more support for early phase ideation and problem formulation to avoid AI product failures. We discuss the implications for future resources aiming to help practitioners in designing AI products.

## CCS CONCEPTS

• **Human-centered computing → Interaction design process and methods**.

## KEYWORDS

human-AI guidelines, human-AI interaction, people AI guidebook

*Work done as an intern researcher at Google Research.
†Work done while at Google Research.

## 1 INTRODUCTION

Artificial intelligence (AI) presents new challenges for designing AI-enabled products and services. AI products[1] can make inference errors that are difficult to anticipate, leading to unpredictable behaviors, breakdowns in user experience (UX) and harm [10]. There has been substantial research in the HCI literature on AI's interaction design challenges, including issues around explainability [25], intelligibility [1], trust [106], user control [62], and setting user expectations [53]. Studies report that practitioners who do not have a background in AI or data science (e.g. designers, product managers, domain experts) struggle to understand what AI can do to effectively engage AI product development [24, 78]. It remains challenging to envision and prototype AI-enabled products and experiences; and to anticipate and mitigate AI's UX breakdowns and potential harm [40, 104].

In response, practitioner-facing resources from large technology companies (e.g. Microsoft [4], Google [75], Apple [5], IBM [42]) became available to aid some of these challenges. These are mostly in the form of guidelines and design patterns that offer best practices around specific design considerations, such as explainability, user control, and feedback. While there are more than a dozen human-AI interaction design guidelines [7], little is known about how these guidelines are *actually* used in practice.

In this paper, we investigated how practitioners working on AI product teams use the People + AI Guidebook (the guidebook) [75] as an instance of industry human-AI guidelines that is relatively understudied [100]. Our focus was not on assessing whether guidelines are effective in improving the user experience of products [59] or assessing how industry guidelines compare to each other [100]. Instead, we sought to understand the felt experience of practitioners and their perceptions around human-AI guidelines. We hoped that their reflections would reveal new insights around the challenges practitioners face and inform the development of future AI design guidelines and resources.

We interviewed 31 practitioners (i.e. UX researchers, UX designers, design managers, product managers) across 23 AI product teams from 14 organizations. We asked practitioners what led them to turn to guidelines, and how they employed AI design guidance in their practice. We probed if and how guidelines help, as well as remaining challenges and their needs for additional support.

---

[1]We use the term AI products to refer to products or product features where AI capabilities enable and impact the human experience.

Our study revealed several interesting insights:

(1) Practitioners not only use guidelines for addressing AI's design challenges, but also for education; for developing internal resources; for cross-functional alignment; and for buy-in within their teams and organizations.
(2) While guidelines are useful for problem solving, practitioners desire better support for problem framing, ideation, and the selection of the right human-AI problem.
(3) We observed an emergent, hybrid design process reconciling user-centric and AI-centric approaches, where participants worked to match AI capabilities with human needs.

This paper makes two contributions. First, we provide a rare description of how human-AI guidelines are actually used in practice. We detail the impact of these resources on practitioners' roles and practices. Second, we highlight opportunities for AI guidelines and resources to better address practitioners' needs, specifically for early phase support. We foreground that AI products fail when they fail to address real human needs or wants. We report emergent practices for discovering value and risks, and discuss implications for future research aiming to develop resources for supporting human-centered approaches to AI.

## 2 RELATED WORK

### 2.1 Challenges of Designing AI Products

Advances in AI have enabled a new set of capabilities, opening up a new design space for interactive systems. However, designing AI-enabled interactions brings unique design challenges. Even simple AI applications (e.g. two-class classifiers) can make inference errors, leading to UX breakdowns and harm [49]. A significant amount of research investigated the challenges of designing human-AI interaction, such as intelligibility [1], explainability [60, 98], user control [62, 87], feedback [89], error recovery [41], and setting user expectations [47, 53]. Parallel to these efforts, a growing body of work has focused on uncovering and mitigating inherent biases and potential harm of AI applications for developing fairer systems [39, 57, 64, 94, 99]. Despite these efforts, it still remains difficult to anticipate the potential errors and harm a not-yet deployed AI system might cause [104].

Another, complementary strand of research has investigated the practices and roles of industry practitioners who work on cross-functional teams. Researchers report that practitioners without a background in AI (e.g., designers, product managers, domain experts) have challenges in engaging data and AI [24, 78]. Product professionals struggle to understand what AI can do, and cannot easily formulate business problems into data science problems [54, 65, 72, 77, 78, 108]. From an interaction design perspective, AI presents unique challenges for ideating and rapid prototyping novel, implementable AI experiences [24, 90, 104] (see [104] for a review of AI's design challenges). Designers mostly join projects towards the end for communicating the AI system's output to end users, and are rarely involved in problem formulation [21, 24, 105].

In addition to challenges specific for individual roles, there are many challenges in cross-disciplinary collaboration, mainly due to a lack of shared workflow, shared language, and shared expertise

[31, 50, 78]. Some researchers characterized this as a gap between AI and product expertise [102, 105]: data science and AI teams struggle to elicit user requirements from product teams, and they tend to overlook how the AI system will generate value for users [54, 72, 78]. Product teams (e.g., PMs, designers) tend to envision AI concepts that cannot be built or too high stakes to deploy an imperfect AI model [24, 104]. Recent research investigating the high failure rate of AI initiatives shows that most AI product failures stem from miscommunications and misalignment between team members, especially in early phase problem formulation [26, 46, 95].

A small but growing number of studies explored the emergent best practices for designing AI products. Successful practitioners formed close collaborations between data science and product teams [54, 90, 103, 105], where product teams worked to sensitize data science and AI partners to human-centered approaches [107]. Design teams in particular served as facilitators on AI teams to establish a shared understanding and alignment between different disciplines using interim artifacts and visualizations [105, 107]. Several researchers highlighted that such artifacts and boundary objects [56, 88] might play a key role in scaffolding cross-disciplinary collaboration among AI practitioners and stakeholders [15, 90, 102, 105].

In terms of data and AI literacy, Yang et al. [103] interviewed designers who successfully engaged *AI as a design material*. These participants only had a high level understanding of AI (e.g. what a model and label is), and they worked with abstractions of AI capabilities and value propositions (e.g. predicting user intent to surface relevant actions). Another study found that design teams created internal AI resources, such as an AI capability matrix and a collection of examples, for educating themselves and engaging their clients in early phase ideation and problem formulation [105]. Similarly, studies on data science and AI practitioners revealed that they worked to sensitize their cross-functional partners, such as PMs and domain experts, to AI concepts [54, 72, 78]. The way they worked resembled design practice [54]; they probed their collaborators to get at the underlying problem, to frame and reframe it as a well-defined data science problem [54, 72, 76]. Finally, research showed that effective AI teams had an approach similar to startups [83], where they would focus on a minimum viable product at the intersection of what is feasible with AI models and data, and what is desirable and viable for customers and users [96, 103, 105, 107].

Current research efforts to aid practitioners in designing human-AI interaction broadly fall into four themes. First, researchers created resources to improve non-technical stakeholders' literacy in data science and AI concepts [37, 52], however there is no consensus on what constitutes a *good enough* technical understanding [104]. Second, researchers developed prototyping tools and design-oriented data exploration workflows with little or no code to allow first-hand experimentation with data and AI [17, 27]. Third, researchers created resources, guidelines, and design patterns to provide best practices for addressing AI's design challenges [4, 5, 42, 75]. Fourth, researchers explored data and AI through design-led inquiry to share first-person accounts, as well as emergent solutions such as new methods, boundary objects and metaphors [8, 15, 23]. Reflecting on these, some researchers argued that AI might require design processes beyond user-centered design; for instance, processes such as matchmaking [11] that seek for potential value in existing AI systems and datasets [20, 29, 101, 107].

## 2.2 Human-AI Interaction Guidelines

HCI literature has a rich history of providing design principles to improve the user experience and usability of interactive systems, typically in the form of guidelines [6, 9, 41, 67, 87], checklists [44, 81], and design patterns [12, 18, 92]. Prior work investigating the application of guidelines in practice showed that practitioners often consider guidelines useful yet face challenges in selecting, prioritizing, or translating them for their work [19, 71, 91].

Recently, building on the large body of research on interaction with intelligent systems, technology companies have proposed principles, guidelines, and design patterns for human-AI interaction to aid product professionals in designing AI products (e.g., [4, 5, 42, 75]). These resources cover a comprehensive list of design considerations, such as explainability (e.g. "make clear why the system did what it did" [4]), control (e.g. "give people familiar, easy ways to make corrections" [5]), and feedback (e.g. "let users give feedback" [75]). In addition to general guidelines, some resources focus on specific interactions (e.g. voice interactions [3], chatbots [68]) or considerations (e.g. ethics [43]) (see a review in [7]).

While there are more than a dozen human-AI guidelines, little work studied how practitioners use these guidelines and resources. Prior research focused on validating the effectiveness of guidelines in improving the user experience [59], or studying guidelines comparatively to reveal the landscape of considerations [45, 100]. A study mapping existing guidelines and resources to current product development processes pointed out that AI guidelines mostly help with late phase efforts (e.g. prototyping), and that resources for early phase are scarce (e.g. ideation) [104]. Recently, a few researchers investigated how specific resources and guidelines are used by industry practitioners, such as fairness toolkits [22] and conversational UI guidelines [51]. In the same spirit, this work investigates if and how human-AI guidelines in [75] are used, which we briefly introduce below.

## 2.3 People + AI Guidebook

Developed by Google, the People + AI Guidebook (the guidebook) is "a set of methods, best practices, and examples for designing with AI" [75]. The content is arranged around five sections: (1) chapters outlining design considerations (i.e. User Needs + Defining Success; Data Collection + Evaluation; Mental Models; Explainability + Trust; Feedback + Control; Errors + Graceful Failure), (2) design patterns with sensitizing examples demonstrating patterns and anti-patterns (e.g. "emphasize how the app will benefit users, avoid emphasizing the underlying technology"), (3) case studies of real-world AI products, (4) a workshop kit with a facilitator guide, (5) glossary of AI related terms. The guidebook employs three hypothetical app examples for design patterns – a running app, a plant classification app, and a learning app – to illustrate how principles might be applied in practice.

The guidebook is similar to other industry guidelines (e.g., [4, 5]) in that it contains design patterns and guidelines for human-AI interaction. Yet it has more content compared to other resources, including the chapters that provide an overview of fundamental AI concepts such as precision-recall trade-offs and confusion matrices [100]. We were interested in exploring the use of the guidebook – both in form and content– as a relatively more comprehensive yet understudied resource.

## 3 METHOD

### 3.1 Study Design

We wanted to understand how practitioners on AI product teams actually use the guidebook in their practice. We broadly defined "practitioners" as people who work on a team developing AI-enabled products and services. We aimed to recruit broadly across many roles (e.g., designers, product managers, data scientists, engineers, domain experts, etc), with the hope that our target audience would emerge through our recruitment process. Our inclusion criteria included having experience with the guidebook concerning AI/ML projects.

We recruited participants using a mix of purposive and snowball sampling [38]. We contacted an initial set of participants through our personal connections with industry practitioners. We then asked them to share any relevant contacts to grow our set of participants. In parallel, we recruited through online platforms (e.g., Linkedin, Twitter) using a brief screener form to help us target suitable participants. We emailed direct contacts across over 20 companies, asking them to disseminate our recruitment message within their networks. Whenever possible, we tried to recruit people in different roles from the same team to gain additional perspectives.

We paid attention to three aspects during recruitment. First, we tried to recruit participants from a broad range of technology areas. Second, we tried to include organizations differing in size and service type (i.e., consumer or enterprise). Third, we wanted to learn how the guidebook is used both as an internal and external resource. Therefore, we tried to have a balanced sample of participants within and beyond the organization where the guidebook was developed (16 out of 31 participants from Google).

We conducted semi-structured interviews with 31 practitioners across 23 AI product teams from 14 companies (9 large companies, 4 startups, 1 non profit). While we did not limit participation to any role, our participants mainly included designers (e.g., design managers, user experience designers, user experience researchers) and product managers. Although we had participants who shared that their colleagues from engineering and data science roles used the guidebook, we were not able to recruit any participants from those roles. Participants worked on teams developing AI products across a wide range of technology areas, including recommender systems, medical diagnosis, text prediction, and more. Table 1 provides a summary of the participants' teams, roles, and technology areas.

Our interview protocol had three main parts. First, we asked participants about their roles, practices, and workflows on AI projects to gain background information and context. Next, we asked them about the use of the guidebook. We asked them to walk us through a recent case where they incorporated the guidance in their work. We probed them about how they discovered the guidebook, and whether and how the guidebook helped with specific challenges. Finally, we asked them about the remaining challenges that were not covered by the guidebook, their needs for support, and any other areas for improvement. Whenever possible, we encouraged participants to share any artifacts created as part of using the guidebook or as aids for relevant challenges. These included worksheets, design patterns and examples, case studies, and workshop materials. All interviews were conducted remotely on a video conference platform.

**Table 1: Participants' teams and technology areas. Roles included product manager (PM), design manager (DM), user experience designer (UXD), and user experience researcher (UXR).**

| Team | Technology Area | Roles | ID | Size | Team | Technology Area | Roles | ID | Size |
|---|---|---|---|---|---|---|---|---|---|
| 1 | Business analytics | UXD | P20 | <100 | 13 | Recommender system | UXD | P14 | 100-1,000 |
| 2 | Business analytics | DM, PM, UXD | P8, P10, P11 | >10,000 | 14 | Recommender system | DM | P28 | <10,000 |
| 3 | Code completion | DM, UXD, UXR | P4, P9, P26 | >10,000 | 15 | Recommender system | UXR | P31 | <10,000 |
| 4 | Code completion | PM | P16 | <100 | 16 | Resume screening | DM | P29 | >10,000 |
| 5 | Conversational AI | UXD | P21 | <10,000 | 17 | Search & retrieval | UXD | P17 | >10,000 |
| 6 | Financial forecasting | UXD, DM, UXD | P1, P2, P15 | >10,000 | 18 | Search & retrieval | UXD | P19 | <10,000 |
| 7 | Fraud detection | UXD, DM | P18, P23 | >10,000 | 19 | Search & retrieval | PM | P22 | >10,000 |
| 8 | Image classification | DM | P7 | >10,000 | 20 | Speech recognition | UXR | P13 | <10,000 |
| 9 | Medical diagnosis | UXD | P5 | >10,000 | 21 | Text prediction | UXD | P6 | >10,000 |
| 10 | Medical diagnosis | UXD | P30 | >10,000 | 22 | Text prediction | UXD | P27 | >10,000 |
| 11 | Personal healthcare | DM | P24 | <100 | 23 | Warranty processing | UXD | P25 | <100 |
| 12 | Recommender system | UXR, UXD | P3, P12 | >10,000 | | | | | |

## 3.2 Data Analysis

We recorded and transcribed the interviews, and documented the artifacts participants shared during or after the interviews. Each interview lasted between 30 to 60 minutes, resulting in 18.5 hours of audio in total. We analyzed the transcripts using affinity diagramming as standard methodology from contextual design [48, 66]. We followed a bottom-up process, pulling out insights and generating codes for participant utterances. We also analyzed the artifacts, such as customized guidelines, patterns, and self-created resources, with an eye for latent needs. We then iteratively reviewed and synthesized these into high-level themes related to current practices, challenges, needs for support, and emergent approaches. We present these in detail in the Findings section.

## 3.3 Positionality

We acknowledge that our own experiences and positionality shape our perspectives and approaches. Our academic backgrounds draw on a mix of disciplines in our research, including AI, HCI, and design. Our professional experience primarily involves working in industry and academia on projects involving AI practitioners as collaborators. Additionally, two of the authors were involved in the development of industry human-AI guidelines in their prior work [75], while they did not conduct interviews and data collection.

## 4 FINDINGS AND DISCUSSION

We present our findings around three themes: using the guidebook as a means for building a culture around human-centered AI, practices for incorporating AI guidance in products, and emergent practices for remaining challenges. Within each high-level theme, we present sub-themes detailing specific challenges and the corresponding use of the guidebook for support. We share implications and design opportunities at the end of each high-level theme. The themes are not exclusive; some aspects spread across themes.

Related to participants' team roles, we noticed a distinction between lead roles (i.e. design managers, product managers) and individual contributor roles (i.e. UX designers, UX researchers) regarding their challenges in developing AI products. While all participants operationalized the guidebook in some way, lead roles described it as more of a strategic resource whereas individual contributor roles described it as a tactical resource. In the detailed findings below, we note where this difference was evident. Interestingly, we did not observe any differences regarding internal and external use (within or outside Google).

## 4.1 Establishing a Human-Centered AI Culture

When we asked participants about their experiences with the guidebook, we mainly expected to hear how they used it to inform the design of AI products. Interestingly, participants' answers revealed that they often used the guidebook as a means for building a culture around human-centered AI within their teams and organizations. In this section, we detail the use cases around (1) education; (2) developing internal resources; (3) cross-functional alignment; and (4) gaining credibility and buy-in. We discuss the implications for the design of future human-AI guidelines at the end of the section.

*4.1.1 Participants used the guidebook for educating themselves, their teams, and organizations.* Most participants reported that the literacy around data and AI is low among PMs and UX practitioners within their teams and organizations. Especially, participants in lead roles reported ongoing educational efforts to address this key concern. For example, several participants (P8, P13, P14, P21, P22, P31) gave talks and presentations to large audiences using the guidebook content to create awareness on AI's design considerations: *"I gave an internal talk maybe to 60-80 designers and PMs with a lot of lessons from the people AI guidebook." (P21)* A few participants (P17, P28, P31) mentioned using the guidebook chapters as a skeleton to create an internal course for product teams –typically for PMs and designers– who are new to working with AI: *"We have recently created an ML for designers course internally as part of educating our peers ... [We used] examples that we either have shipped or we just experimented with and haven't shipped to tell the story of how to work with AI and ML." (P28)*

We also observed that participants who did not have any prior experience working with AI (P3, P4, P12, P14, P15, P18, P25) used the guidebook for self-education: *"My first objective was: I need to get smarter about this topic. ... [after digesting the guidebook* content] *Then my next step was, how do I communicate this to my PM, engineering and UX team?" (P3)* Several participants found the guidebook approachable as an entry point: *"I went from knowing nothing to being someone who has a substantive amount of knowledge in the UX of ML." (P4)* All participants shared that they became proponents of the guidebook, educating their teams through sharing specific chapters or patterns or through short talks and presentations. Similarly, a majority of participants reported learning about the guidebook through their colleagues or broader professional connections. A few participants mentioned becoming aware of it through industry conferences or discovering it organically when searching for resources.

### 4.1.2 The guidebook provided a benchmark for developing internal resources and guidelines.
Many participants (11 teams out of 23) revealed that they have self-created AI design resources and guidelines for internal use, and that they referred to the guidebook as a benchmark when they were developing their own resources. Among these, some participants noted that they benchmarked multiple resources, including industry guidelines (e.g. [4, 5, 42, 79]) and academic papers (P21, P28, P29, P30). Few participants (P4, P23) also shared that they have plans to develop internal resources, and are *"using this as a source of truth in the meantime." (P23)* Additionally, within large technology companies, we observed multiple internal resources and guidelines that were customized for a product area (e.g. healthcare AI design guidelines).

Internal resources often took the form of playbooks, documentation hubs, and slide decks, and included domain-specific patterns and examples [detailed in section 4.2.4]. A design lead mentioned using the guidebook design patterns when developing a design system and component library for AI-based products: *"We referenced several of the patterns when creating design components specification." (P29)* For example, they created components for representing intelligent agents, such as chatbots, in a way that sets user expectations and mental models.

When probed about the need to create domain- or organization-specific resources, participants highlighted the importance of relevance to own organizational context and domain: *"There is a strong need to appropriate it to your own domain, to your own industry, and to your own organization as you probably have very different needs. … Regardless of how good the guidebook or these kinds of tools could be, we would have to do that appropriation step and pick and choose what is relevant for us." (P30)* Interestingly, several participants noted that even with the custom developed internal resources, creating awareness within their organizations remained a challenge.

### 4.1.3 Participants used the guidebook for easing the challenges of cross-functional collaboration.
Participants brought up several challenges around cross-functional collaboration in AI-based projects, and they described how they used the guidebook to overcome such challenges. For example, several participants mentioned establishing a shared language as a major benefit: *"Establishing the vocabulary around the space is one of the biggest values that I see in an artifact like this. Because sometimes we use the same terms and don't refer to the same thing, or use different terms for the same thing." (P20)* Some participants noted that it empowered them in participating discussions and product decisions: *"Anytime we work from a principles based approach, it levels the playing field." (P4)*

Participants in lead roles spoke of using the guidebook to sensitize their engineering and data science teams to AI's design considerations, and to human-centered AI in general:

> *"[We did a workshop using the guidebook chapters] focusing on how do we make these features more human centered … The engineering and the data science team found it incredibly useful because it helps them to better understand everything you may have to think about when designing for ML. … That was very beneficial for building the human centered muscle within the team, regardless of the actual outputs of the exercises." (P8)*

Participants shared that these efforts helped to facilitate a shared understanding and alignment within their teams, and were well received by cross-functional partners.

### 4.1.4 The guidebook was used for gaining credibility and buy-in for design recommendations.
Several participants spoke of using the guidebook as an artifact for negotiation and alignment on design recommendations (P3, P6, P10, P20, P25, P26, P31). As a PM working on business analytics put it, *"Without the principles that guide the conversation, it can be everyone's personal preference or opinion on how a feature should be or not … What's a stronger pitch or recommendation is, we should do this, because it's tied to this guideline or principle." (P10)* They shared that being able to reference such a resource brought credibility and helped to convince their teams, as the guidance showed tried and true solutions synthesized across many products and academic studies. Interestingly, all UX researchers (P3, P13, P26, P31) described using the guidebook as a literature review; they cited the relevant guidance when presenting research findings: *"[When we shared our interview study findings] data scientists' first reaction was, you make all these claims and recommendations, how generalizable are these? It was really nice to be able to say, we looked at the literature and the people AI guidebook, and brought these things together." (P26)*

### 4.1.5 Discussion and Implications.

- **Support learning within AI design resources and guidelines.** While human-AI guidelines are primarily designed to support practitioners in designing interactions, our findings indicate that practitioners also use these resources for self and organizational education. Future AI design resources might be designed to better support bi-directional learning: human-centered AI concepts for data science and engineering teams, and AI concepts as it relates to product development for product teams. For example, resources can be presented in various forms, such as short videos or interactive modules for self-learning, or as slides for delivering talks and presentations. In contrast to the abundance of educational resources for technical AI/ML concepts (e.g. ML crash-courses [33, 73]), there is little educational content for human-centered AI ([69, 82] as rare examples). Future resources can be served in a course-like format to support both individual and organizational learning.

- **Support cross-functional communication and collaboration.** A large body of HCI research has suggested creating boundary objects to effectively scaffold collaboration between AI practitioners [15, 102]. Our findings offer preliminary evidence highlighting the potential of AI design resources to facilitate the collaboration between team members and stakeholders from different disciplines. We see a great opportunity for future resources to be explicitly designed for use by multiple teams and roles across product development. Future research should investigate the specific needs of different roles to better understand how to support team members with different knowledge and expertise.

- **Support adaptation and appropriation of resources for the development of domain- and organization-specific guidance.** Prior research discussed the inherent trade-off between generality and specialization for developing AI resources and guidelines, noting that specific applications might require specialized guidelines [4]. Our results confirm this; teams and organizations have a strong desire to develop their own resources or adapt existing ones to contextualize the guidance for their specific application domains. To this end, resources and guidelines from industry and academia served as a benchmark for practitioners. One clear implication is to design resources for adaptation and appropriation by providing explicit affordances. For instance, resources might contain guiding questions to help practitioners assess which design considerations and guidelines might be most relevant to their domain, service type or AI technology areas. Similarly, guidelines can include diverse case studies and examples demonstrating how a particular design consideration might apply to different contexts and interaction scenarios.

## 4.2 Incorporating AI Guidance in Products

In this section, we present how practitioners addressed some challenges of designing AI products by incorporating concrete guidance from the guidebook. We detail how practitioners operationalized the guidebook, and their needs for enhanced guidance.

### 4.2.1 Framing Human-AI Problems.
Participants often commented that AI projects are heavily driven by technical considerations, and that a human-centered approach can be missing. A major challenge in AI-centric development was ensuring that the AI product was addressing a real user need. Several participants brought up the chapter on defining user needs (chapter 1) and stressed the importance of problem framing for setting up AI projects for success. A PM working on search and recommenders (P22) systems shared: *"I use the guidebook for problem framing to make sure that machine learning is solving an actual people problem. … [Before] it was being treated as a technical solution, there was no thought into what kind of user problem is being solved. And because of that, a lot of the things that people were shipping didn't work."*

P22 gave an example where the data science team was assigned a "time to last result" metric to predict when all the relevant results have been loaded to stop the search. When the team conducted user research, they realized that it was a nonexistent problem: *"[Users] didn't even care if the results are still loading, they only cared about the first three results. [The data science team] were optimizing for a metric that wasn't even relevant." (P22)*

This was a shared sentiment regardless of whether participants worked at large software companies or startups. Several participants (P3, P13, P18, P19, P22, P27) similarly recalled projects that failed or got canceled, and reflected that they should have spent more time framing and defining the problem: *"We did some concepts and tested [the feature] with users, no one really wanted it. … In retrospect, I would have pushed back [on the idea early on]." (P19)* In some cases, participants were able to reframe the problem to iterate and pivot to a solution grounded in user needs: *"Is there a different experience we can create? Maybe it's a useful model. Working with data science, you can often pivot and do something else." (P19)*

Participants described using the workshop kit and worksheets in the guidebook for conducting problem framing workshops where they collectively defined user needs and success metrics with their cross-functional team members. A UX designer working on medical diagnosis shared an activity where they asked their team to separately fill out a worksheet detailing the intended user, the intended use, and the value for the user: *"There was actually divergence … Then it becomes a discussion to align on these." (P5)* Some participants shared that the guidance helped them assess whether AI was the right fit for a problem or whether they should use heuristic or rule based approaches instead (P3, P12, P17, P29, P31). However, some participants felt limited in questioning problem frames as their role centered around providing AI solutions: *"I'm definitely not a person who believes machine learning should be used for everything, but a lot of the projects that I work on are pre-baked with machine learning. So I don't often get to ask 'should we actually use ML systems?'" (P21)*

### 4.2.2 Addressing AI's specific design considerations.
Participants often used the guidebook to identify and address AI's design challenges around setting expectations and mental models, providing explanations, designing feedback and control mechanisms, and mitigating errors (chapters 3-6). Some participants noted that they had no prior knowledge around these challenges: *"A huge part of it was figuring out, what are the design considerations that we need to have when using machine learning?" (P19)* Most participants recalled cases where they used the guidance to influence product decisions: *"[Based on] the chapters in the guidebook, we provided a lot of feedback and explanations, and a lot of control to users." (P29)* Specifically, they mentioned examining user journeys and service blueprints of their products to identify where critical issues, such as trust issues or errors, might arise. A majority of participants shared conducting focused workshops and design sprints to collectively envision solutions that operationalize specific guidance (e.g. explainability).

Most participants became aware of problematic issues and UX breakdowns during testing or after launch. Indicators of breakdown were typically lack of user adoption, lack of trust or satisfaction, and error reports from users. Similar to reports in AI fairness literature [39], several participants commented that addressing design challenges *"tend to be reactive despite our best efforts" (P6)*, and expressed a desire to anticipate these challenges early on: *"the majority of our learnings happened in internal beta testing when the feature is live, where it's time expensive and resource heavy to make any changes. … [Designing AI products] has been a learning curve for our team at every stage of the development cycle, from preparing for launch to post-launch to internationalizing." (P13)*

Other participants reported that while they were aware of AI's design considerations, it takes time and several iterations to craft and implement thoughtful solutions: *"We launched some of our initial [image classification] features without really good feedback loops … We've had to see the real user impact of some of the features in order to really put these principles into practice. So the guidebook was helpful in getting from nowhere to somewhere. We're seeing even now how much more we need to do, how we need to build controls that are more flexible, more granular or have better clarity to help people truly avoid harmful experiences."* (P7)

### 4.2.3 Explainability and trust was the most referred chapter; data collection and evaluation was the least referred chapter.

While participants referred to all the aforementioned chapters, explainability and trust was the challenge they spoke of the most. Interestingly, we observed that participants used "trust" as a broad term to describe many problematic situations. Some participants spoke of intelligibility, where users did not understand what the AI system could do and would not use the suggested system actions (P1, P3, P9, P18). Others described usability and user acceptance issues, where users understood what the system does, but would not trust it to do what it claimed it was capable of doing (P13, P14, P24, P25). Few participants mentioned cases where the system produced biased or incorrect outputs when describing lack of user trust (P6, P7, P29). Several participants reflected on the complexity of trust as a concept: *"[part of the challenge] is thinking of the different dimensions of trust, as it's an open-ended and complex concept. It's almost like saying, what is love? It means different things to different people in different situations."* (P6)

We also noticed that only a few participants referred to the chapter around data collection and evaluation (chapter 2). These participants spoke of issues around bias in machine translation (P6), automatic speech recognition (P21), resume screening (P29). When explicitly probed, most participants shared that they did not use the guidance around data as they worked with pretrained models. Among the participants who were individual contributors (20 out of 31), only three were involved in the collection of initial training data. Instead, participants mostly worked on collecting user feedback data after launch (chapter 5). Only one participant, a design lead on business intelligence applications, mentioned using the guidance on data collection for designing systems for data labelers who annotate training data (i.e. *'design for labelers and labeling'*) (P29).

### 4.2.4 Patterns and examples work well, but there is a need for domain-specific examples.

Nearly all participants referenced the design patterns and examples, whether for developing internal resources or for addressing specific design challenges as individual contributors. Several participants commented that the patterns made the overall content and guidance actionable and practical. The major challenge with design patterns was the relevance of examples for participants' own domain. Participants' criticisms included the lack of breadth; narrow focus on consumer-based examples; and limited use of real-life examples.

A majority of participants shared that they collected their own domain-specific design patterns and examples. When we asked participants how they curated these patterns and examples, few dimensions seemed important. First, participants wanted patterns that are hyper relevant: *"… it's not just medical, but specifically examples around medical imaging."* (P5) Similarly, a UX designer pointed out that best practices and patterns might differ within a specific technology area, such as language interactions: *"Confidence in voice is very different than confidence in chat. In voice, confirming what the user said would be the best practice; for chat it's surfacing multiple options."* (P21) Second, participants searched horizontally, looking at adjacent domains to see a breadth of examples. For example, P13 working on mobile communication referred to AI applications in email; and P24 working on personal healthcare devices looked at personalized home energy products for patterns around automation. Third, relevance to own service type – consumer vs enterprise applications – stood out as a salient aspect in collected examples.

In addition to collecting existing examples, several participants mentioned creating hypothetical concepts and UI mocks for contextualizing how a pattern might apply to their product. They often referred to the guidebook's pattern representation illustrating patterns and anti-patterns as *industry standard*, noting they adopted this structure when creating their internal patterns. A UX lead (P7) shared an internal example where they introduced even more granularity in the pattern and anti-pattern spectrum. For example, they sketched four UI mocks for collecting user feedback: one without any feedback mechanism, one with feedback mechanism without any explanations, one with a generic explanation, and one with a better explanation. Internal pattern collections became a part of internal resources [detailed in section 4.1.2] and were typically documented in slide decks or visual workspaces (e.g. Figma [28]). However, several participants desired more effective mechanisms for sharing and reuse of this knowledge with other teams.

### 4.2.5 Participants observed a positive impact on products, but shared many challenges around measuring and evaluating impact.

A large body of work in HCI has demonstrated the effectiveness of the guidelines (e.g., [40, 53, 55, 59, 61, 93, 106]). Similarly, our participants often observed a positive impact on products, and reported overall positive perceptions around the usefulness and effectiveness of the guidebook. However, several participants expressed challenges around measuring and evaluating the impact of guidelines on their products (P2, P6, P11, P13, P25, P29). Participants mostly spoke of impact through metrics such as increased adoption rate, satisfaction, engagement, task completion rate, and resolution time, measured through qualitative or quantitative studies. However, they desired better support for measuring the impact of incorporating AI design guidance in their products, including metrics and methods: *"It would be great to hear what others have done to understand how to measure success or understand what is 'good enough' for launch."* (P13)

Related to measuring impact, several participants emphasized that it was critical to communicate the product impact and the business case to the larger product team for the guidelines to be prioritized. A UX researcher (P3) spoke of constraints of the product development process: *"The only push back I got from my team was, what priority is this? Should this be part of the MVP (minimum viable product)?"* They shared that the design team was able to make a case for incorporating explanations by creating a few design concepts and validating the concepts through a user evaluation study.

### 4.2.6 Discussion and Implications.

- **Provide context-specific design patterns and guidance.** Prior research has emphasized the need for contextualization in AI resources and toolkits for fairness [22, 97]. Our findings echo this; participants worked to contextualize the design considerations and patterns to their products and applications. Future resources and guidelines should provide patterns and examples that (i) are **domain-specific** (e.g., best practices for providing explanations in healthcare); (ii) include **a breadth of service types** (e.g. consumer vs enterprise applications); and (iii) display **a breadth of patterns** for specific AI design considerations (e.g. different ways to collect explicit feedback). Future work is needed to explore ways to index and organize such a pattern library where practitioners can contribute their work. What would be a good level of abstraction for contributing unpublished or hypothetical examples remains an open question.

- **Provide both patterns and anti-patterns with varying degrees of granularity.** Traditionally, HCI research has followed well-defined formats for design patterns, such as Alexandrian and Tidwell forms [2, 18, 92]. Interestingly, a majority of participants in our study seemed to have adopted the pattern and anti-pattern structure that is commonly used in industry design resources [32, 35] when developing their internal patterns. Studying how practitioners create and appropriate AI design patterns marks a clear space for HCI and design research. Future work can investigate these forms through artifact analysis to identify which dimensions are critical for practitioners to effectively understand and apply AI design patterns "in the wild".

- **Provide guidance on measuring and evaluating the impact of human-AI guidelines.** HCI literature has provided several methods for evaluating design principles' impact on usability and user experience, such as comparative usability testing and heuristic evaluation [16, 74]. However, applying these methods for measuring the impact of AI guidelines can be costly and complex [59]. Our findings confirm this challenge; many participants expressed difficulties in measuring the impact of guidelines and principles in the context of their specific applications. Future resources should provide explicit guidance on measuring and evaluating the impact, including relevant metrics and methods for pre-launch (e.g. lab studies) and after launch (e.g. real-world data). For instance, recent research suggested using factorial surveys for evaluating AI systems that do not yet exist [59]. Guidance on evaluating the impact may help practitioners in getting organizational buy-in and championing human-centered AI practices forward.

## 4.3 Emergent Practices for Broader Challenges

In this section, we present additional challenges that were not supported by the guidebook, and detail the emergent practices we observed in response to these challenges.

### 4.3.1 Practitioners need more support for early phase AI ideation, problem selection and formulation.
A major critique from participants was that while the guidelines were helpful for solving existing problems, little help was provided for finding the right problems: *"The guidebook really focuses on optimizing a design you already have … like diagnosing why something might not be working versus the envisioning step." (P13)* Several participants highlighted needs for support in the early stages of the product development for ideation and problem formulation with cross-functional partners. As a PM put it (P22), *"There is even a step before [defining user needs]: what is a problem that you could apply machine learning for? … [Product managers] don't even know how to ask a data scientist whether we could solve this through machine learning."* They emphasized the reciprocal relationship between product development and model development, as they worked to find use cases where machine intelligence can improve products and user experience:

> *"Given a current solution or an envisioned future solution, how do we ideate about what AI could be doing or what the data requirements would be? … That probably happens before you even get to such a guidebook." (P30)*

Interestingly, there were few participants who created their own resources and artifacts for scaffolding ideation and problem formulation with AI. Similar to reports in prior literature [103, 105], these were AI capability abstractions that delineated a subset of relevant capabilities and example applications. For example, a UX designer working on business analytics (P1) shared an "AI pillars" framework they created with their data science team prior to conducting a visioning workshop. The framework involved the capabilities *discovery, recommendation, prediction,* and *automation,* where each capability was detailed with the types of inferences and value (e.g. *discover seasonality and industry trends to help businesses understand market and customers*). A UX researcher (P13) shared a similar framework outlining AI capabilities (e.g. *identify, inform, anticipate,* etc) and example applications illustrating the value proposition (e.g. *save time, save effort, provide assistance, reduce distractions,* etc). Reflecting on how this capability framework helped with systematic brainstorming, P13 noted:

> *"The framework offered [designers and PMs] a way to think about what ML could do, whereas beforehand, it was just the ML team who understood what it could do. There wasn't this shared understanding of what [ML] could offer. … Before, people were throwing out ideas [sporadically]. Now [our] team has a hyper focused mindset of 'what are all the things we could do?' to brainstorm [product] features within that capability."*

Finally, a few participants talked about the prioritization, assessment and selection of ideas as challenges (P13, P19, P22, P29). A design lead shared a matrix detailing the performance (AI model's confidence and the quality of data), the risk of errors (false positives and false negatives), and the user's familiarity with the system: *"When working with some healthcare implementations [that were] high risk with a low confidence level and where the user is not very familiar with the system, we provided lots of feedback, explanations and controls to users." (P29)* We observed several assessment matrices across different teams, including a modified impact-effort matrix [36] that captured impact, feasibility, and risk; and a two-by-two matrix capturing risk (high or low) and frequency of use (one-off use or everyday use).

### 4.3.2 Practitioners described a hybrid design process to reconcile user-centric and AI-centric approaches.
When talking about the broader challenges of designing AI products, several participants expressed tensions between user-centered design (UCD) and AI product development. Some participants reflected on the fact that having an already built AI model limits the solution space: *"We were already coming at it from, where can we add intelligence to add value? Not necessarily from a blank slate like, what pain point should we solve?" (P3)* Other participants recalled cases where the UCD process alone was not effective in identifying AI opportunities. For example, a UX researcher tried engaging end users in AI product development, but simply asking users where they need intelligence did not work: *"How do we develop a method that isn't just asking people where they want machine intelligence? Instead we can learn what's painful about their process … [with an eye] out for where we can fit [intelligence] in." (P26)*

Additionally, there were challenges in brainstorming without thinking about limitations and constraints of data and AI: *"[In visioning workshops] teams tend to think of ideas that are too far ahead, then you have challenges in grounding [ML ideas]." (P17)* Some design leads and PMs simply recognized this as a shortcoming of UCD: *"[For] the ML or AI side, we don't really have a process baked in at all. How do we discover a new solution for customers or a feature or product that might be a good fit for an AI application?" (P23)* Interestingly, several participants described a hybrid process where they started both with user needs *and* AI models or data, and sought to explore good matches in this problem-solution space. As a PM put it, this approach helped discover both value and downstream risk (P16):

> *"[We do a lot of sketching] who are these personas, why [is this] useful? What are the range of interactions that could happen? [Then thinking about] all the different inputs for training the model that may have downstream value … [not only to ensure relevance but also] to minimize the downstream risk."*

### 4.3.3 Discussion and Implications.

- **Broaden the scope of human-AI resources and guidelines to provide support for early phase ideation and problem formulation.** Recent literature on human-centered AI and fairness in AI have highlighted the importance of early phase ideation, problem formulation and selection for identifying the right problem to solve [77, 85, 104, 105]. Our findings confirm prior speculations on the lack of guidance and support for early phase AI product development [104]. In response, some researchers explored curating AI capability abstractions and example applications to sensitize product teams to what AI can do [101]. Our findings echo this approach; several participants in our study expressed a need to understand what AI can do for discovering problems; and a few participants in our study shared similar AI capability frameworks they had created as aids. Future resources could include taxonomies and frameworks delineating what AI can do, and what types of problems are best suited for AI to better support practitioners in ideation and problem formulation.

- **Support practitioners in incorporating resources into their existing processes and workflows.** Similar to findings from prior literature [24, 90, 96, 103], our findings point out that practitioners start AI product development in different stages of AI model development; and that the entry point might not be apparent for every team or product. Future resources can better support practitioners in incorporating guidelines and design considerations into their existing processes by explicitly connecting them to different stages in AI product development. For instance, resources around data collection can be used in upstream product development for concept generation; and in downstream product development for assessing current data collection practices. Similarly, providing actionable forms and methods, such as design workshops and patterns, may help practitioners in operationalizing design guidance. In particular, workshop resources that can be adapted to collaborative work (e.g. visual workspaces such as Miro [70] and Figma [28]); and to time constraints (e.g. one hour vs one day long workshop kits) can lower the barrier to entry for practitioners.

## 5 OPEN QUESTIONS & FUTURE DIRECTIONS

### 5.1 Discovering the Right Human-AI Problem

HCI researchers have cleverly distinguished problem setting from problem solving [14, 84]. Our community has specialized methods for *sketching* – exploring the right thing to build; and *prototyping* – building the thing right. Our study revealed that current AI guidelines are helpful in evaluating and refining human-AI interactions. However, there is a need for generative resources for discovering and formulating problems where AI might be a good solution.

Challenges in problem framing are common in cross-functional software development [30, 58], yet working with AI capabilities adds additional barriers. AI products and experiences are difficult to sketch: once a dataset is collected and a model is built, the design space becomes limited to certain problem framings [104]. Participants in our study recognized this challenge, and expressed needs for support in early phase ideation, problem selection and formulation. Similar to prior work [103, 105], we observed self-created AI capability frameworks that explicitly detail what AI can do and how it might create value to broadly explore the design space. We also uncovered assessment matrices detailing model performance and risk for prioritizing and selecting ideas. We hope that the emergent resources presented in this work will inspire future resources to support practitioners in discovering the right human-AI problem.

On a higher level, our findings point to a more consequential problem: AI products fail when human centered perspectives are lacking, especially in the early problem formulation phase. Several participants shared hard learned lessons through project failures, highlighting how problem framing is essential for success yet overlooked. This problem seems remarkably similar to the early days of software development, before human-centered design became the norm for reducing the risk of developing unwanted technology [14]. Despite recent evidence [26, 46, 95], AI product failure is rarely discussed in the HCI literature. Most HCI research around AI is focused on usability; yet our study shows that practitioners need more support for assessing usefulness. While the lack of methods for usefulness is not specific to working with data and AI [34], identifying the right thing to design becomes more critical for AI products as it can be extremely difficult and costly to make changes once an AI system is built. We see a real opportunity for HCI practitioners to play a critical role in ensuring that AI products are solving real problems. Future research should provide methods and resources for iteratively framing, reframing, and pivoting to ensure a match between AI solutions and human needs.

## 5.2 Reconciling User-centric and AI-centric Innovation Processes

Previous work surfaced the tensions between user-centered design process and the AI development process, and speculated that AI product development may require design processes beyond UCD [29, 104, 107]. Our findings echo this: we observed emergent processes that seemed to be a hybrid between user-centric and AI-centric processes. Several teams shared that during problem formulation and ideation, they limited their solution space to existing AI models and data sets. Similarly, teams spoke of their process as customer or feature discovery, where they conducted needfinding studies with an eye for particular AI solutions. As prior research noted [103], their process resembled technology-centric product development (e.g. matchmaking [11]); yet it differed in that participants were not considering all possible customers and users. Instead, they were looking at a small set of users related to their products and services.

Future research should further investigate these new, hybrid design processes for reconciling user-centric and AI-centric approaches. Similar to [77], ethnographic studies exploring different modes of AI product development across different teams, organizations, and product settings might reveal insights into how product ideas are selected, defined, and prioritized during early problem formulation. New knowledge into AI product innovation processes may open up new opportunities for embedding human-centered perspectives throughout the AI development lifecycle. We encourage HCI researchers to explore, speculate, and formalize these emergent processes by studying them in the wild.

## 5.3 Guidance around AI Fairness

Our study revealed parallels between AI's design challenges and AI's fairness challenges. Similar to prior literature [22, 39, 63], practitioners in our study expressed difficulties in mitigating potential bias (P6, P7, P16, P20, P21, P29, P31). While industry AI guidelines provide some guidance around fairness and bias, these tend to be high level and underdeveloped [100] (e.g. "mitigate social biases" [4], "consider bias in the data collection and evaluation process" [75]). Our findings show that practitioners currently lack a vocabulary to describe fairness issues. They need more granular guidance detailing risks around trust and fairness. How might fairness and bias issues manifest themselves in specific domains and AI applications? Madaio et al.'s work detailing practitioners' fairness desiderata provides a great starting point [64]. Future research should explore how to contextualize and communicate fairness considerations in a way that is easy to incorporate into current practices (e.g. design patterns, case studies, workshop kits).

Recent discussions in AI fairness literature have highlighted the importance of problem selection and formulation for bringing harm minimization closer to the early stage of technology design rather than considering risks post-deployment [13, 77, 86]. Our findings echo this approach. Most of the participants in this study worked with already trained models, yet participants that were involved in the selection and collection of data for training models seemed to have more awareness of potential bias and harm. They spoke of concurrent model and product development where they sketched product ideas and interactions early on to envision downstream value and risk, which in turn informed data and model work. We suspect that these emergent practices around broad exploration of the problem-solution space provide a great opportunity for identifying potential fairness risks early in the development process.

From a broader perspective, our study highlights an overlapping scope between human-AI guidelines and AI fairness toolkits. Currently, AI fairness toolkits are typically targeted towards engineering and data science roles who drive the upstream AI development processes (e.g. data collection, model building). However, our study and recent literature indicate that many other product roles (e.g. product managers, UX designers, software engineers, domain experts) inform data collection practices and contribute to improving fairness efforts [39, 64, 80]. Future research should investigate how these seemingly separate guidelines overlap, and how to scaffold fairness efforts across multiple team roles and stakeholders.

## 6 LIMITATIONS

Our study has several limitations. First, this study focused on a single resource, the People + AI Guidebook, as an instance of human-AI guidelines from large technology companies. Future research should explore the broader landscape of AI design guidance, including resources from industry, academia, and more. Second, by recruiting participants who already have used the guidebook, we may have sampled practitioners who find AI guidelines useful and are unusually motivated to use such resources. Third, our participants mainly included designers and product managers, there are many other roles involved in AI product development (e.g. software engineers, data scientists, domain stakeholders) whose perspectives were not covered in this work. Future work should take a broader recruitment approach to recruit participants from a wider range of roles and organizations, including people outside of academia or industry (e.g. government agencies, civil society organizations). Finally, our findings are based on retrospective interviews that were conducted over a limited period of time. Future research should include longitudinal ethnographic studies to fully understand the uses and impact of human-AI guidelines and resources.

## 7 CONCLUSION

In this study, we conducted an empirical exploration of how industry practitioners use the People + AI Guidebook to gain nuanced insights into how human-AI guidelines are employed in practice. Through our interviews with designers and product managers on AI product teams, we found that AI guidance is useful not only for addressing AI's design challenges, but also for establishing a culture around human-centered AI. We discovered that practitioners need more support in early phase problem framing and ideation. We documented their emergent practices and self-created resources to make up for lack of support. We hope that the design implications we uncovered will inform the development of future resources and guidance for designing AI products.

## ACKNOWLEDGMENTS

We are grateful to Michelle Carney, Gabe Clapper, Di Dang, Kristie Fisher, Shay Gray, Jess Holbrook, Patrick Gage Kelley, Maysam Moussalem, Kristen Olson and Rebecca Salois for their support; Elizabeth Churchill, Clara Kliman-Silver, Tiffany Knearem and Dan Russell for their feedback on the manuscript. We thank internal and external study participants for their invaluable input. We also thank the Center for Responsible AI and Human Centered Technology at Google Research for enabling this work.

This work was conducted by the People + AI Research team, funded by Google Research. The authors declare no additional sources of funding. The legal department of Google participated in the review and approval of the manuscript; and the decision to submit the manuscript for publication. Aside from the authors and their collaborators, Google had no role in the design and conduct of the study; access and collection of data; analysis and interpretation of data; or preparation of the manuscript. The authors declare no other financial interests.